# Forward and backward mapping of image to 2D vector field using fiber bundle color space

**A.A. Snarskii, I.V. Bezsudnov**

National Technical University of Ukraine "Igor Sikorsky Kyiv Polytechnic Institute" (NTUU KPI), Kiev, Ukraine

Corresponding author: A.Snarskii, e-mail: asnarskii@gmail.com

## Abstract

We introduce the concept of a fiber bundle color space, which acts according to the psychophysiological rules of trichromacy perception of colors by a human. The image resides in the fiber bundle base space and the fiber color space contains color vectors. Further we propose the decomposition of color vectors into spectral and achromatic parts. A homomorphism of a color image and constructed two-dimensional vector field is demonstrated that allows us to apply well-known advanced methods of vector analysis to a color image, i.e. ultimately give new numerical characteristics of the image. Appropriate image to vector field forward mapping is constructed. The proposed backward mapping algorithm converts a two-dimensional vector field to color image. The type of image filter is described using sequential forward and backward mapping algorithms. An example of the color image formation on the base of two-dimensional magnetic vector field scattered by a typical pipe line defect is given.

## 1. Introduction

The concept of color is formed by the psychophysiological peculiarities of human perception of electromagnetic waves in the visible range, the so-called human trichromatic perception of colors. The color for a human is associated not only with the spectrum of electromagnetic radiation entering the eye, but also with its processing sequentially by the eye, retina and finally by corresponding parts of the brain, see for instance [1-4]. There is no one-to-one correspondence between wavelength and color perception. The yellow color, for example, appears both when the eye is irradiated with monochrome light (550 - 590 nm), and in the case when red and green light act simultaneously.

There are a number of established facts describing the psychophysiological patterns of color perception and also a number of concepts describing color perception [5,6]. They are based on the idea of primary colors and color models (for example, RGB or CMYK color models) [7-9], on the concept of achromatic (black, gray or white) color, on the empirical rules how to add or subtract colors and manage their intensities, etc.

Various planar graphical representations of colors [1-6] are employed where different plane points refer to different colors. Probably the first one was Maxwell's triangle [10,11], nowadays the more general so-called shark fin is widely used [1-8]. Three-dimensional representations use to specify different colors as vectors in 3D space with the axes corresponding to the primary colors, for instance, red, green and blue.

Another option is to represent and to manipulate the colors within the image as a quaternion [12-17], where the quaternions correspond to the primary color's intensities. There are other representations as well [18-23].

The main purpose of all those conceptions is to make it possible to determine which color will be felt by a human eye by measuring different physical characteristics of incident light, i.e. spectral composition, intensities of the primary colors and/or the intensity of the achromatic color, etc.

Of course, in all kinds of representations, the basic rules of human trichromatic color perceptions have to be fulfilled. Below we partially summarize them close to [1], the same or similar are indicated in other monographs [1-8]:

• Each wavelength corresponds to a single color,

• A mixture of two wavelengths produces a perception of a single color lying in the spectrum between them, excluding the edges of the spectrum, any spectral color can be reproduced with a set of spectral colors,

• There are pairs of colors that give a perception of white (achromatic) color,

• For any color (except green) there is a so-called complementary one which, when mixed with the original, will give an achromatic color,

• Any color can be reproduced by mixing an achromatic color with a pair of monochromatic colors of a specific wavelength.

In this paper, to describe a human color perception we use such a mathematical construction as a fiber bundle [24-26]. The base space of the bundle is a given color image - a two-dimensional space where the intensities of three primary colors are set for every point, a fiber space of the bundle is a three-dimensional vector space with color vectors that are determined by the intensities of the primary colors of the original image in base space.

The first part of the article describes the main properties of constructed fiber bundle color space and color vectors and shows that the patterns of psychophysiological color perception are fulfilled. In the second part, using the fiber bundle approach we demonstrate that any color image can be associated (mapped to) with a vector field. In the third part we propose an backward (inverse) mapping of any two-dimensional vector field to the image that can be used for convenient data presentation for the investigator or be a part of proposed image filter.

## 2. A vector field in color space

Further we will use as primary colors those used in light sources investigations i.e. red, green and blue.

We introduce a fiber bundle: $(\mu, CS \otimes CP, CP)$, $\mu : CS \to CP$, where $CP$ is the base space of fiber bundle, further we will call it a Color Plane ($CP$), i.e., the plane with the image, which is color image. Intensities of primary colors are $I_R$, $I_G$ and $I_B$ (values in $CP$ are marked by capitals). As a rule, the image is a set of color pixels on a plain regular lattice but in common case, it could be the set of three functions. The fiber space ($CS$) is the three dimensional space – the color space having orthogonal axes $Or$, $Og$ and $Ob$ with unit vectors $\mathbf{e}_r$, $\mathbf{e}_g$ and $\mathbf{e}_b$ (small letters correspond to the $CS$ values). The mapping $CS \to CP$ is denoted by $\mu$. Fig.1 shows necessary axes and vectors in both $CS$ and $CP$ spaces.

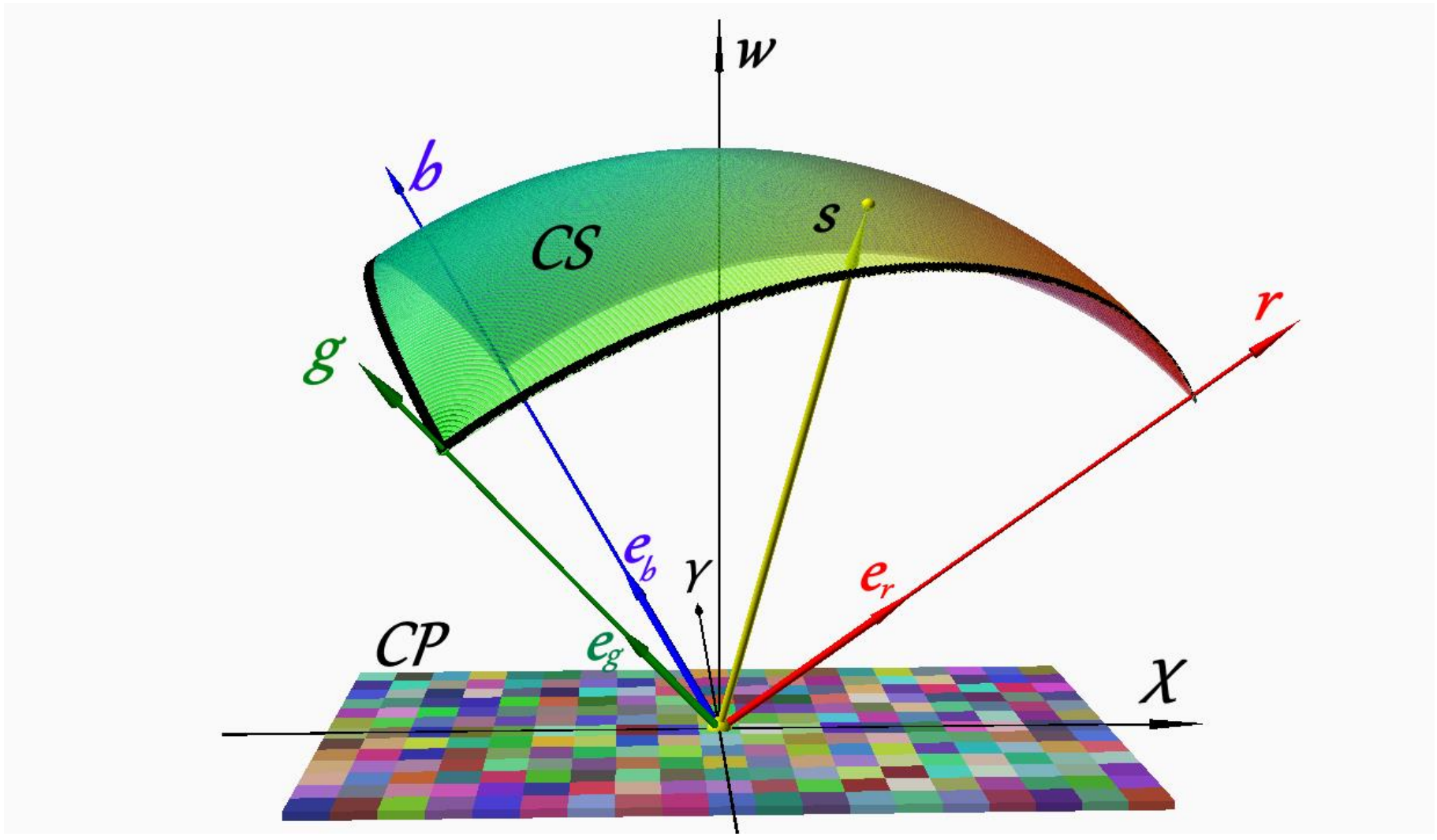

Fig. 1 Fiber bundle color space. $CP$ - color plane (base space), $CS$ - color space (fiber).

The color vector in $CS$ is defined by its projections $r$, $g$ and $b$:

$$\mathbf{s} = r\mathbf{e}_r + g\mathbf{e}_g + b\mathbf{e}_b \tag{1}$$

where $r = \sqrt{I_R}$, $g = \sqrt{I_G}$, $b = \sqrt{I_B}$, therefore $r$, $g$, $b$ are positive and the length of the color vector corresponds to pixel brightness in $CP$: $I = \mathbf{s}^2 = r^2 + g^2 + b^2 = I_R + I_G + I_B$.

Note that the brightness of each color is given in normalized units: equal intensities of primary colors $I_R = I_G = I_B$ results in achromatic color (mentioned usually as gray or white if intensities reach its maximum or even black if they are zeros). In such a unit, the color vector directed along the spatial diagonal of coordinate planes (vertical $Ow$ axis on Fig.1) corresponds to the achromatic color.

Fig.1 demonstrates the part of the sphere surface of radius $|\mathbf{s}|$ bounded by $CS$ coordinate planes $rOg$, $gOb$, $bOr$. Points of that surface have equal brightness $I = |\mathbf{s}|^2$ but different colors. All the possible spectral colors are on that surface and completely cover it. This spherical triangle is the

analogue to Maxwell's color triangle [10]. Triangle corner points correspond to the selected primary colors, and binding lines (parts of large circles) correspond to monochromatic colors residing between primary colors, e.g., yellow between red and green, etc.

Suppose arbitrary color vector $\mathbf{s}$ has minimal blue intensity, I.e., $b = \min(r, g, b) > 0$ . In this case the color vector $\mathbf{s}$ can be decomposed into the sum of two vectors: spectral $\mathbf{s}_{spect}$ and achromatic $\mathbf{s}_w$.

$$\mathbf{s} = r\mathbf{e}_r + \mathrm{g}\,\mathbf{e}_g + b\mathbf{e}_b = (r-b)\mathbf{e}_r + (\mathrm{g}-b)\mathbf{e}_g + b(\mathbf{e}_r + \mathbf{e}_g + \mathbf{e}_b) \equiv \mathbf{s}_{spect} + \mathbf{s}_w, \tag{2}$$

$$\mathbf{s}_{spect} = (r-b)\,\mathbf{e}_r + (\mathrm{g}-b)\,\mathbf{e}_g, \;\; \mathbf{s}_w = b(\mathbf{e}_r + \mathbf{e}_g + \mathbf{e}_b), \tag{3}$$

since $\mathbf{e}_r + \mathbf{e}_g + \mathbf{e}_b \,||\, Ow$.

In the above example the spectral vector $\mathbf{s}_{spect}$ belongs to the $rOg$ plane and achromatic vector $\mathbf{s}_w$ directed along the $Ow$ axis (see Fig.2a). All components of $\mathbf{s}_{spect}$ and the value $|\mathbf{s}_w| = \sqrt{3} \cdot b$ are also nonnegative. In the case $g = \min(r, g, b) > 0$ the spectral vector $\mathbf{s}_{spect}$ goes to $rOb$ plane, but achromatic vector $\mathbf{s}_w$ is always directed along the $Ow$ axis, etc. Such a decomposition can be performed for any color vector and always $\mathbf{s}_w$ is along the $Ow$ axis.

## 3. The 2D vector field for the image

Above color vectors allow us to create the 2D vector field for the given image.

Actually, the vector field can be composed as the orthogonal projection of vectors $\mathbf{s}$ in $CS$ fiber on the base $CP$ plane with $XOY$ Cartesian orthogonal coordinate system (see Fig. 1). Notice, that the point $O$ resides on $Ow$ axis and also it implements the coordinate origin in both $CS$ and $CP$. Therefore, the projection vector $\mathbf{S}$ (see Fig. 2b) in $CP$ will be as follows

$$\mathbf{S} = S_X \mathbf{e}_X + S_Y \mathbf{e}_Y \tag{4}$$

In the proposed case, the projection of the achromatic component $\mathbf{s}_w$ of the color vector $\mathbf{s}$ is zero (see Fig. 2a) and does not contribute to the vector field in $CP$, although the set of values $|\mathbf{s}_w|$ forms a separate scalar field of the achromatic component, that can be referred to as achromatic image.

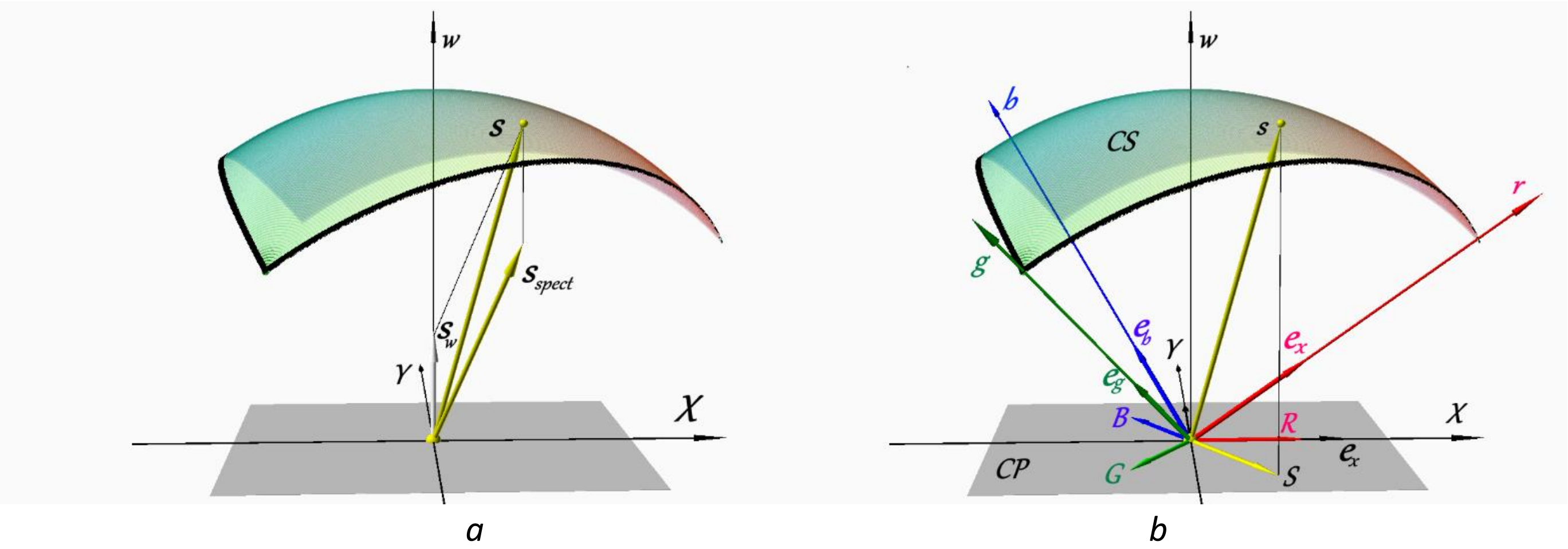

Fig. 2. Color vector and its projection: a) color vector $\mathbf{s}$ in the fiber color space $CS$ as a sum of spectral and achromatic vectors , b) projection to the base $CP$ plane.

Here we set the direction of the $OX$ axis in $CP$ along the projection of $\mathbf{e}_r$ see Fig. 2b. The projection $\mathbf{S}$ is formed only by the spectral vector $\mathbf{s}_{spect}$.

Using the notation of Figs. 1 and 2, analytical relationship for $\mu : CS \to CP$ is given below. We denote as $\mathbf{R}$, $\mathbf{G}$, $\mathbf{B}$ the projections of $\mathbf{e}_r$, $\mathbf{e}_g$, $\mathbf{e}_b$. The mapping relations are

$$\begin{aligned}
\mathbf{R} &= \sqrt{2/3}\cdot\mathbf{e}_X \\
\mathbf{B} &= -\sqrt{1/6}\cdot\mathbf{e}_X + \sqrt{1/2}\cdot\mathbf{e}_Y \\
\mathbf{G} &= -\sqrt{1/6}\cdot\mathbf{e}_X - \sqrt{1/2}\cdot\mathbf{e}_Y
\end{aligned} \tag{5}$$

The color vectors $\mathbf{s}$ in $CS$ are defined as in (1) i.e. $\mathbf{s}=\sqrt{I_R}\mathbf{e}_r+\sqrt{I_G}\mathbf{e}_g+\sqrt{I_B}\mathbf{e}_b$ and the projections will form the vector field on $CP$ acc. to

$$\begin{aligned}
\mathbf{S} = \mathbf{e}_X &\cdot\left(\sqrt{2/3I_R}-\sqrt{1/6I_G}-\sqrt{1/6I_B}\right) \\
+\mathbf{e}_Y &\cdot\left(-\sqrt{1/2I_G}+\sqrt{1/2I_B}\right)
\end{aligned} \tag{6}$$

Fig. 3a shows an image - a still life with a vase and apples with a resolution of 50x50 (the original is taken from [27]), the mapping $\mu$ is applied to this image and the corresponding vector field is constructed see Fig.3b.

Now it becomes possible to apply various well-known vector analysis methods to the two-dimensional vector field $\mathbf{S}$ produced by the original image. For example, to apply the divergence operator $div(\mathbf{S})$ Fig. 3d or $curl(\mathbf{S})$, the latter for a two-dimensional field forms the vectors directed perpendicular to the plane and therefore can be presented as a scalar field also, see Fig. 3e.

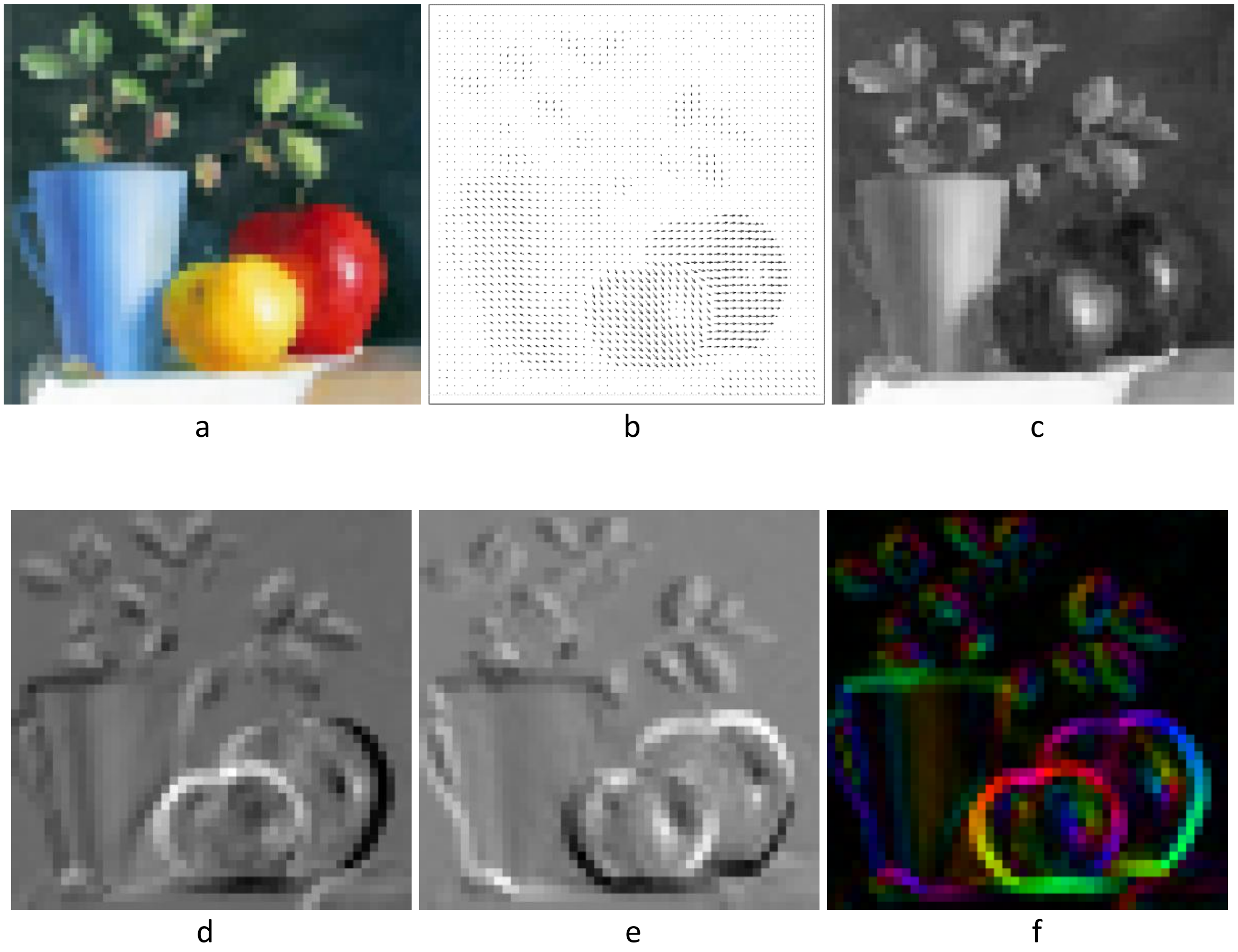

Fig. 3. The vector field and images produced by original picture: a) the original image (color online) is the still life with a blue vase and yellow and red apples presented at 50x50 resolution, b) the constructed vector field $\mathbf{S}$, c) the achromatic scalar field $\mathbf{s}_w$, d) $div(\mathbf{S})$, e) $curl(\mathbf{S})$, $z$ component of it, (for c-e white corresponds maximum to the, and black to minimum values), f) the image composed using d) and e).

### 4. The image built by 2D vector field, visualization

Converting a 2D vector field to an image, i.e. the mapping $v: CP \rightarrow CS$ becomes possible if one maps vectors from $CP$ to the set of spectral vectors $\mathbf{s}_{spect}$ in $CS$, i.e. to vectors located on $CS$ coordinate planes. Such a mapping takes into account that $\mathbf{s}_w$ directed along to $Ow$ axis (see Fig.2) and the projection of any $\mathbf{s}_w$ is zero, in other words we assume the achromatic vector $\mathbf{s}_w = 0$. In this case one can find that every vector in $CP$ corresponds to the vector in $CS$ i.e. vector field in $CP$ produce the set of vectors $\mathbf{s}_{spect}$ in $CS$ and therefore compose a color image.

We will use relations that can be called as the backward or inverse of (5) and (6), taking into account the fact that there is no achromatic component $\mathbf{s}_w$. Let a vector $\mathbf{S} = p\mathbf{e}_x + q\mathbf{e}_y$, where $p,q$ – scalar coefficients, then the mapping $\nu$ can be written in the form of a system of equations for coordinates $(r,g,b)$ of the vector $\mathbf{s}$ in $CS$ (7), allowing one to restore whole image

$$\begin{aligned} & r-b = \frac{p\sqrt{3}+q}{\sqrt{2}}, \\ & g-b = \sqrt{2}\cdot q, \\ & \min(r,g,b) = 0. \end{aligned} \tag{7}$$

The system of equations (7) always has a unique solution because of $\min(r,g,b) = 0$. In other words, equations (7) formularize the idea how to find $(r,g,b)$. If one imagines the perpendicular line to the $CP$ plane, one finds single intersection point with every coordinate plane in $CS$, but correct one is always above $CP$ plane and, consequently, for that point one of the color coordinates is always zero depending on which plane is crossed.

The following are examples of applying a mapping $\nu$ to a vector field.

### 4.1 Image filter using achromatic component

As the first test example, the mapping $\nu$ is applied to the vector field that is formed from the image by applying $\mu$ mapping. Fig. 4a shows the original image of Fig. 3a in high resolution and the image obtained by successively applied mappings $\mu$ and $\nu$. The resulting picture contains only pure spectral colors of various intensities Fig.4b and is completely devoid of the achromatic component. (Of course, removing the achromatic component can be done directly by just subtracting $\min(r,g,b)$ from $r$, $g$, and $b$, but that is just revealing example.)

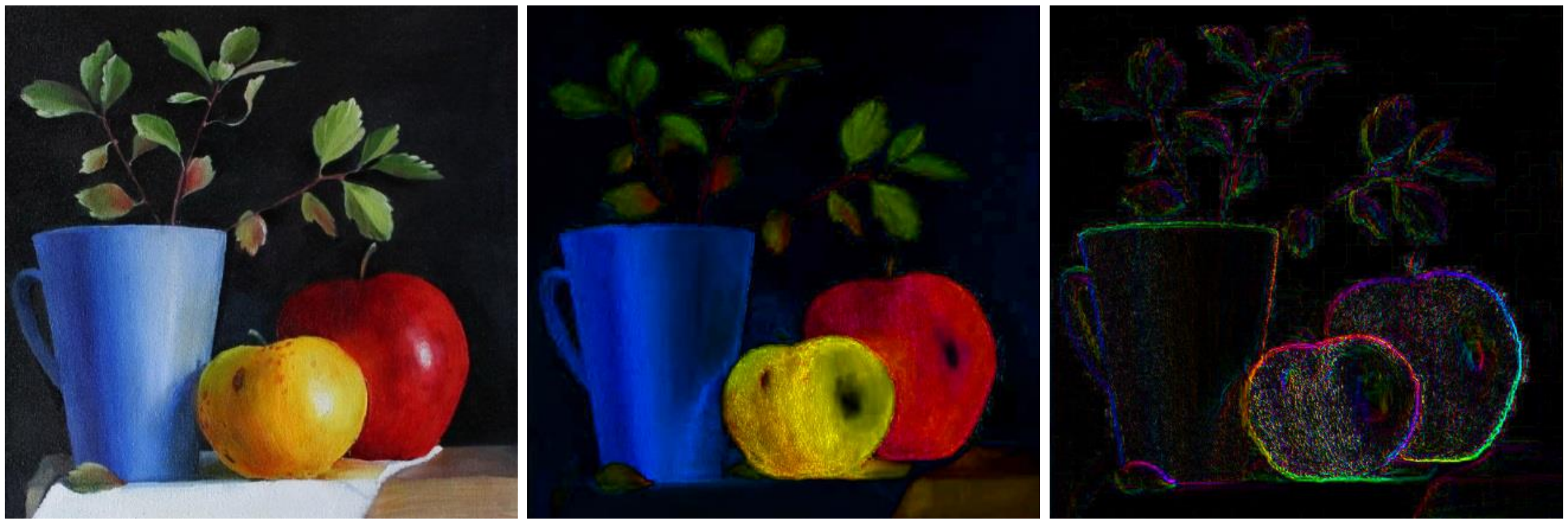

a b c

Fig 4. Application of $\nu$ mapping (color online): a) the original image, b) sequential application of mappings $\mu$ and $\nu$ results in the image without the achromatic component, c) image constructed from gradient of achromatic component of a) using $\nu$.

Thus, the sequential application of mappings $\mu$ and $\nu$ is an adaptive filter for images that removes the achromatic component from the original image. In the proposed form (7), this filter uses the data of only one point. However, it is possible to use a similar procedure that takes into account additional information about the original image, for example, partially use the achromatic component $|\mathbf{s}_w|$.

Another example of usage the achromatic component is the image reconstructed from gradient vector field built from achromatic component of original image is shown on Fig 4c. It should be noted that colors in Fig 4c have no connection to similar-looking colors in the original image Fig 4a. Now one can see not only areas with high values of gradient but conveniently find the direction of high rate of data change.

### 4.2 Image composition using 2D vector field

The second example that could be of practical interest, since it allows us to match a color image to an arbitrary two-dimensional vector field. Consider, for example, the problem of magnetic flux leakage (MFL) that is used to determinate the defects in pipes material of oil and gas pipelines [28,29]. A special inspection device moving inside the pipe generates the magnetic field along the pipe wall, which is scattered on various defects, for example, corrosion pits. The magnetic field scattered by defects is recorded by magnetic field sensors located on the same device composing 2D vector field. Here we present the result of simulating of magnetic field scattering by a complex three pits defect see Fig 5a. The resulting vector field is shown in Fig. 5b.

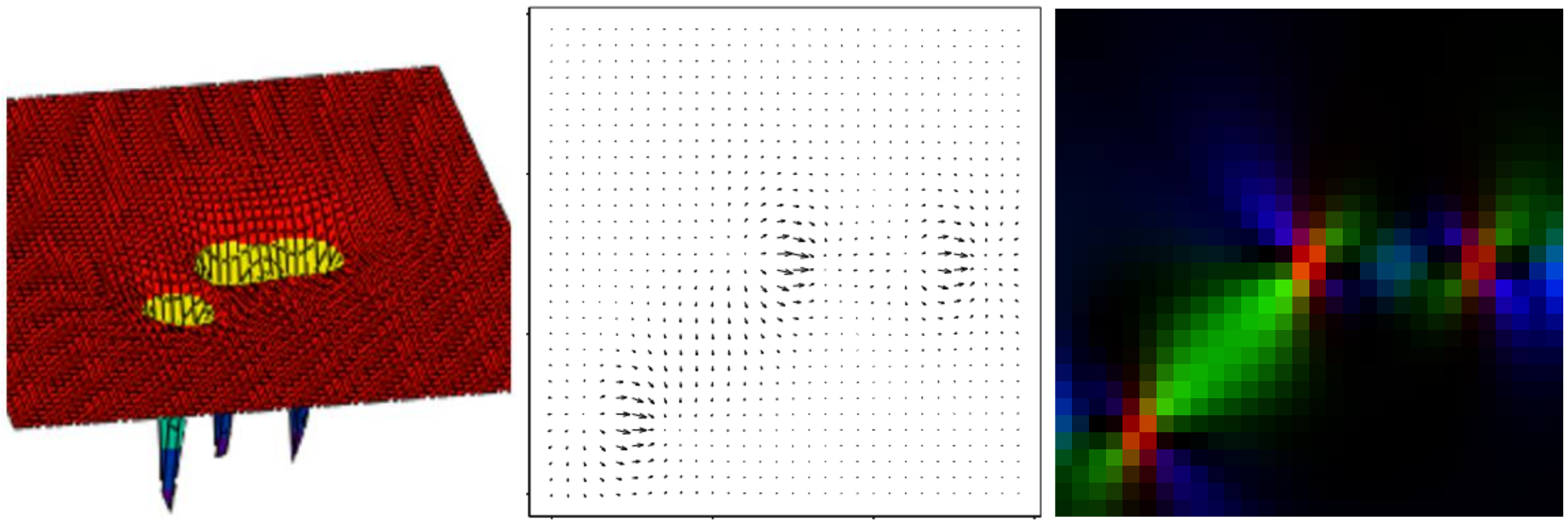

a b c

Fig. 5. MFL defects. a) The sketch of three corrosion pits (defects) on the metal surface, b) vector field of defects, c) a color image produced by the vector field.

Fig. 5c shows the color field corresponding to pits two-dimensional vector field Fig.5b. The places of defects are well distinguished on it. In flaw detection, on the basis of the obtained scattered fields, the inverse problem is solved - the reconstruction of the shape of the defect and its basic dimensions. One of the problems with the MFL method is the need to isolate hazardous areas of the pipeline from a huge dataset of field scattering (24 or more sensors along the surface of the pipe, taking data every millimeter or less at a pipe length of about 100 km). The responsibility of the task of flaw detection at such facilities is very high, therefore in addition to the automatic identification of dangerous places, preliminary "manual check" of data (vector field) is performed. As it can be seen from the above figures, viewing a color image instead of a vector field can be really more convenient and consequently more productive.

The proposed transformation of an arbitrary vector field into an image allows us to evaluate vector fields of large size or simultaneously display a pair (and with appropriate modification of the method even more) of different scalar fields related to the same object. An example of two scalar fields related to the same object can be found in Fig.3d and Fig.3e. Fig.3f shows the image composed from those scalar fields.

## 5. Discussion and conclusions

Introduced fiber bundle color space makes it possible to convert or map any color image to the 2D vector field and allows us to apply well-known advanced methods of vector analysis to a color image, i.e. ultimately give new numerical characteristics of the image.

Backward (inverse) mapping of vector field to color image i.e. visualization of 2D vector field creates images from two-dimensional vector fields using the principles of human trichromacy color vision.

On the basis of the described mappings, the image filter has been proposed that allows us to process a color image in a way to exclude completely or partially the achromatic component, and this can be done adaptively for each point of the image or taking into account other additional information about the image during the transformation.

The concept of fiber bundle color space in future will allow us to determine the distance between two colors and represent the change in the color of a selected point over time as a trajectory in the color space and its further analysis using differential geometry methods. Undoubtedly, it will be also interesting to apply the proposed approaches to working with vector fields of higher dimensions.

**Acknowledgements**

The authors would like to thank Ms. Maria Snarskaya for the opportunity to use her wonderful painting “Last Green Leaves” in this work.